# Energy conservation and reversibility during thermodynamic changes of state in superconductors: Joule heat vs. magnetocaloric cooling

Andreas Schilling* ORCID 0000-0002-3898-2498
Dept. of Physics, University of Zürich, 8057 Zürich, Switzerland

*Corresponding author. *Email address:* schilling@physik.uzh.ch (A. Schilling)

**In the Meissner phase of a superconductor, an external constant magnetic field is shielded by circulating persistent zero-resistance supercurrents that are formed by Cooper pairs. However, a thermodynamic change of state within this phase, such as cooling or heating, inevitably generates normal currents of thermally excited unpaired charge carriers, induced by the time-dependent variations in the local magnetic field. They not only lead to deviations of the magnetic-field distribution from textbook Meissner profiles but also cause dissipative Joule heating. This sharply contradicts the expected reversibility of a truly thermodynamic superconducting state, a fact that has largely been overlooked in the literature. We show that these normal currents also produce a magnetocaloric cooling, which in total instantaneously and precisely compensates for the dissipated heat, thus ensuring overall energy conservation and reversibility. However, the Joule heating and magnetocaloric cooling processes are spatially distinct and should therefore lead to temperature inhomogeneities. We quantify these effects assuming realistic material parameters and conclude that they are challenging to measure with current experimental techniques. Significant temperature gradients are expected only directly at the first-order transition to the superconducting state, where the discontinuous flux expulsion should induce normal currents that are much larger than those deep in the Meissner phase. We also argue that the underlying physics in superconductors is fundamentally identical to that of thermomagnetic generators, where electromechanical work can be extracted from magnetized matter subjected to thermal cycles, and where the magnetocaloric cooling is balanced by the heat supplied from an external thermal reservoir.**

## 1. Introduction

The Meissner effect in superconductors was discovered almost a century ago [1]. According to this effect, an external magnetic field can be expelled by superconducting shielding currents, which, as it has later been shown, consist of Cooper pairs [2]. There is a consensus that at any finite temperature, some of these pairs break up and thereby create unpaired quasiparticles, so that part of the electronic system can then be regarded to behave as normal conducting. This is a central assumption of the well-known "two-fluid model" [3,4]. By applying an external alternating magnetic field, corresponding "normal currents" can be induced which are subject to dissipation (i.e., the generation of Joule heat), a process that is also undisputed and well-studied [5]. Nevertheless, a fundamental aspect has been largely overlooked: such normal currents can also be triggered by induction even in the absence of an external electromagnetic excitation, namely simply by changing the temperature in the superconducting state, since the local magnetic field also varies over time with temperature (see Fig. 1). This raises a fundamental question: how can the existence of dissipative currents be reconciled with the thermodynamic state of superconductivity and with the reversibility of associated thermodynamic changes of state?

## 2. Time dependent London equation

To elucidate the underlying principles at play, we start with the distribution of the local magnetic field $\boldsymbol{B}(\boldsymbol{r})$ in the Meissner state of a superconductor that can be obtained by solving

$$-\Delta \boldsymbol{B}(\boldsymbol{r}) + \frac{\boldsymbol{B}(\boldsymbol{r})}{\lambda^2} = 0, \tag{1a}$$

as it was originally derived from the second London equation in the static limit [6], with solutions adapted to the required boundary conditions. The quantity $\lambda$ represents an effective magnetic penetration depth, which is only in the clean limit and for $\lambda_L \gg \xi$ (where $\lambda_L$ is the London penetration and $\xi$ the coherence length) identical to $\lambda_L$ associated with the density $n_s$ of superconducting charge carriers. In the opposite limit $\lambda_L < \xi$ and in the dirty limit, $\lambda$ is better described by expressions provided by Pippard, which consider non-local corrections and impurities [7] and lead to $\lambda > \lambda_L$, but the general form of equation (1a) remains virtually unchanged. With increasing temperature $T$ towards the critical temperature $T_c$, $n_s$ decreases due to the thermal breaking of Cooper pairs, leading to a rapid divergence of $\lambda(T)$ in all cases (Fig. 1b). At finite temperature below $T_c$, the electronic system can therefore be thought to be composed of a superconducting ($n_s$) and a normal conducting ($n_n$) component, which add up to the total density $n$ of charge carriers. This concept has been very successful in describing the superconducting state [3,4], especially for alternating currents and magnetic fields [5], with $n_n$ often approximated by $n_n \approx n(T/T_c)^4$ [3].

However, if we allow for a time dependence of $\boldsymbol{B}(\boldsymbol{r},t)$ and of the associated current density $\boldsymbol{j}(\boldsymbol{r},t)$ related to $\boldsymbol{B}$, equation (1a) must be modified [8-10]. The total current density $\boldsymbol{j}$ can then be thought of as being split into a supercurrent $\boldsymbol{j_s}$ related to $n_s$, responsible for the Meissner magnetic-shielding effect, and a normal current $\boldsymbol{j_n}$, which is typically present only upon an external stimulus. The curl operation applied to the quasistatic version of the Maxwell equation $\boldsymbol{\nabla} \times \boldsymbol{B} = \mu_0\, \boldsymbol{j} = \mu_0(\boldsymbol{j_s} + \boldsymbol{j_n})$ (valid for $\boldsymbol{B}(\omega)$ frequencies $\omega \ll c/R$, where $R$ is the size of the superconducting body and c the speed of light [11]), leads to a time-dependent version of equation (1a) [8-10],

$$-\Delta \boldsymbol{B}(\boldsymbol{r},t) + \frac{\boldsymbol{B}(\boldsymbol{r},t)}{\lambda^2} = \mu_0 \boldsymbol{\nabla} \times \boldsymbol{j_n}(\boldsymbol{r},t) = \mu_0 \boldsymbol{\nabla} \times \sigma \boldsymbol{E}(\boldsymbol{r},t) = -\mu_0 \sigma \frac{\partial \boldsymbol{B}(\boldsymbol{r},t)}{\partial t}, \qquad (1b)$$

which has been successfully used to solve several time-dependent problems [12-14]. With the vector potential $\boldsymbol{A}(\boldsymbol{r},t)$ in the London gauge belonging to a solution $\boldsymbol{B}(\boldsymbol{r},t) = \boldsymbol{\nabla} \times \boldsymbol{A}(\boldsymbol{r},t)$ of equation (1b), it has been assumed here that the electric field $\boldsymbol{E} = -\partial \boldsymbol{A}/\partial t$ affects $\boldsymbol{j_n}$ according to Ohm's law $\boldsymbol{j_n} = \sigma \boldsymbol{E}$ with the electrical conductivity $\sigma$, while its influence on $\boldsymbol{j_s}$ is described by the London equation through $\boldsymbol{j_s} = -\boldsymbol{A}/(\mu_0 \lambda^2)$. If the variation of $\boldsymbol{B}$ with time is sufficiently slow in a quasistatic limit, $\sigma$ can be regarded as a real, frequency independent quantity and may near $T_c$ be assumed to be close to the normal-state conductivity $\sigma_n$, but it can still vary with temperature (see Section 4 in the Supplementary material). The choice of the actual value of $\sigma$ does not change the main qualitative conclusions of this article as long as $\sigma > 0$, however. If we now allow for thermodynamic changes of state in the Meissner phase, such as the slow cooling or heating in a constant external magnetic field, the $\boldsymbol{B}(\boldsymbol{r},t)$ becomes time-dependent due to the temperature (and therefore time-) dependence of $\lambda(T)$. Consequently, the magnetic-field and current distributions can deviate from those provided by the "static" London equation (1a). In addition, the normal currents generated inevitably lead to irreversible dissipation, no matter how slow the changes of state are. Legitimate considerations have therefore been made about the reversibility in a truly thermodynamic superconducting state as treated in generally accepted models of superconductivity [15-19]. In principle, an additional dissipative term due to the time-dependent energy gap may also be present [20]. In this work, however, we restrict ourselves to the dissipative currents carried by thermally excited unpaired charge carriers. We are discussing these issues and show that the Joule heating due to the normal currents is exactly compensated by the magnetocaloric cooling produced by the very same currents. However, the Joule heating profile is spatially different from the distribution of the magnetocaloric cooling power, leading to small non-equilibrium temperature gradients during such changes of state. We at first develop the underlying thermodynamics, cross check it with the result from the perspective of electrodynamics and then generalize it to study the consequences for thermodynamic changes of state in the Meissner phase of superconductors. The formalism that we will use here is very general and is not restricted to superconductors. The same underlying physics applies to the operation of thermomagnetic generators., where the thermal cycling of magnetized matter can be used to harvest electromechanical work. In such generators, the role of the Joule heat of the normal currents in superconductors, which compensates for the magnetocaloric cooling power, is taken over by the heat provided by an external heat reservoir, which is ultimately driving the generator.

**3. Power distribution and energy balance**

We first consider very generally the changes $du$ in the local internal energy density $u$ of any reversibly magnetized material with magnetization $\boldsymbol{M}$ in a magnetic field $\boldsymbol{H}$ upon variations in the entropy density $s$ and the magnetic induction $\boldsymbol{B} = \mu_0(\boldsymbol{H} + \boldsymbol{M})$ in terms of the first law of thermodynamics,

$$du = Tds + \boldsymbol{H} \cdot d\boldsymbol{B}. \tag{2}$$

By using $\boldsymbol{H} \cdot d\boldsymbol{B}$ (instead of $\boldsymbol{H} \cdot d\boldsymbol{M}$), we are including in $u$ all contributions related to the magnetic field, i.e., also the magnetic energy density of empty space. To identify the relevant mechanism, we first assume complete thermal insulation of the material from the environment in an adiabatic process, $ds = 0$, so that $du = \boldsymbol{H} \cdot d\boldsymbol{B}$. If any time-dependent variation of $\boldsymbol{B}(t)$ induces extended currents $\boldsymbol{j}_{ind}(\boldsymbol{r}, t)$ in closed conducting paths according to Faraday's law of induction, these currents generate an additional "induced field" $\boldsymbol{H}_{ind}(\boldsymbol{r}, t)$ related to $\boldsymbol{j}_{ind}$ via the Maxwell equation $\boldsymbol{\nabla} \times \boldsymbol{H}_{ind} = \boldsymbol{j}_{ind}$. The total local magnetic field $\boldsymbol{H}(\boldsymbol{r}, t)$ can then be interpreted as the sum of the local magnetic field $\boldsymbol{H}_{\boldsymbol{0}}(\boldsymbol{r}, t)$ that would be present in the absence of induced currents, and $\boldsymbol{H}_{ind}(\boldsymbol{r}, t)$. In an extended but finite conducting body, $\boldsymbol{j}_{ind}$ must be finite at its surface. The corresponding boundary condition for $\boldsymbol{H}_{ind}(\boldsymbol{r}, t)$ is therefore $\boldsymbol{H}_{ind,surf} = \boldsymbol{0}$ so that the total magnetic field remains continuous there. The additional change $du_{ind}$ in the internal energy density is $du_{ind} = \boldsymbol{H_{ind}} \cdot d\boldsymbol{B}$. For a varying $\boldsymbol{B}(\boldsymbol{r}, t)$ we can therefore assign an associated local power density $p_{ind}$ to account for changes of $u_{ind}$ with time,

$$p_{ind}(\boldsymbol{r}, t) = \frac{\partial u_{ind}}{\partial t} = \boldsymbol{H_{ind}}(\boldsymbol{r}, t) \cdot \frac{\partial \boldsymbol{B}(\boldsymbol{r},t)}{\partial t}, \qquad (3)$$

with $p_{ind} = 0$ at the surface. An alternative derivation of equation (3) can be made directly from electrodynamics using Poynting's theorem $\partial u_{ind}/\partial t = -\boldsymbol{\nabla} \cdot \boldsymbol{S} - \boldsymbol{j_{ind}} \cdot \boldsymbol{E}$ [21], applied here specifically to the present problem. With the Poynting vector $\boldsymbol{S} = \boldsymbol{E} \times \boldsymbol{H_{ind}}$ and the vector identity $\boldsymbol{\nabla} \cdot \boldsymbol{S} = (\boldsymbol{\nabla} \times \boldsymbol{E}) \cdot \boldsymbol{H_{ind}} - (\boldsymbol{\nabla} \times \boldsymbol{H_{ind}}) \cdot \boldsymbol{E}$, it yields the same result with $\boldsymbol{\nabla} \times \boldsymbol{E} = -\partial \boldsymbol{B}/\partial t$ and $\boldsymbol{\nabla} \times \boldsymbol{H_{ind}} = \boldsymbol{j_{ind}}$. To properly account for energy conservation, the power density $p_{ind}$ in equation (3) must be considered along with all other sources of power, such as possible internal Joule heat production, or externally supplied heat to achieve non-adiabatic reversible temperature changes in an experiment.

We now allow for such a controlled temperature variation at a rate $dT/dt$ in a constant $\boldsymbol{H_0}$. A non-zero $\partial \boldsymbol{B}/\partial t$ can then originate only from the varying $\partial \boldsymbol{M}/\partial t = (\partial \boldsymbol{M}/\partial T)(dT/dt)$ in a magnetized material with a temperature dependent magnetization $\boldsymbol{M}(T)$. Lenz's law dictates that for any combination of signs of $dT/dt$ and $\partial \boldsymbol{M}/\partial T$, the $\partial \boldsymbol{B}/\partial t$ and $\boldsymbol{H_{ind}}$ produced by the induced currents always have opposite signs, so that strictly $p_{ind} < 0$ (see Supplementary material, Section 1), and we may identify this energy withdrawal from the material by the currents as a magnetocaloric-cooling process [22-25]. At the same time, the induced currents $\boldsymbol{j_{ind}}(\boldsymbol{r}, t)$ themselves may be subject to dissipation. The underlying physical laws responsible for this magnetocaloric process are identical to those at work in the well-known technique of adiabatic cooling, where a thermally isolated substance with temperature and field dependent entropy is exposed to a varying external magnetic field, leading to a change in sample temperature [25]. In fact, the expression found in many textbooks to calculate a differential temperature change $\Delta T$ due to a small adiabatic increase in the magnetic field, $\Delta \boldsymbol{H}$, is $\Delta T = -T/C(\partial \boldsymbol{M}/\partial T)\Delta \boldsymbol{H}$, where $C$ is the heat capacity per volume [25]. This can be formally obtained from Eq. (3) for the case where $\boldsymbol{H_0}$ is constant over time, and using $\partial \boldsymbol{M}/\partial t = (dT/dt)(\partial \boldsymbol{M}/\partial T) = \partial \boldsymbol{B}/\partial t$. Here, $\Delta \boldsymbol{H}$ refers to the additional change in the magnetic field due to the induced normal currents, $\boldsymbol{H_{ind}}$, which in our case always points in opposite direction to $\partial \boldsymbol{B}/\partial t$. The amount of magnetocaloric heat involved is $Q = C\Delta \boldsymbol{T}$, and the associated magnetocaloric power density becomes $p = dQ/dt = -(dT/dt)(\partial \boldsymbol{M}/\partial T)\Delta \boldsymbol{H} = (\partial \boldsymbol{B}/\partial t)\boldsymbol{H_{ind}} = p_{ind} < 0$.

The equation (3) is a very general result and is by no means restricted to superconductors, or even to the mere presence of magnetic materials. If, as in the present case, $\partial \boldsymbol{B}/\partial t$ stems from a substance with non-zero $\partial \boldsymbol{M}/\partial t$ in a constant $\boldsymbol{H_0}$ alone, $p_{ind}(\boldsymbol{r},t)$ corresponds to the local magnetocaloric cooling power, leading to a real local cooling of the material. This is, e.g., the basis for the operation of certain thermomagnetic generators, where a conducting pick-up coil is placed around a paramagnet with a strongly temperature dependent magnetic susceptibility in a closed electrical circuit [26]. Upon repeatedly cycling the temperature, an external surplus of heat must be supplied by a heat reservoir to compensate for the magnetocaloric heat that is converted into electrical energy. In such cyclic experiments, however, the extraction and return of magnetic energy cancel each other out exactly, so that, on average, no magnetic energy is converted into electrical energy. In the opposite limit of $\partial \boldsymbol{B}/\partial t$ related to an external alternating magnetic field $\partial \boldsymbol{H_0}/\partial t$ only and in the absence of magnetized matter, $p_{ind}(\boldsymbol{r},t)$ represents the local energy withdrawal from the electromagnetic field by the induced currents. This is the case for the classic skin effect in metals, or when an empty pick-up coil in a closed resistive circuit draws energy from a varying $\boldsymbol{H_0}(t)$. It is essential to note that the power distribution in such processes is generally spatially inhomogeneous, because the Joule heat is generated near the edges of an experimental set-up, while the energy withdrawal occurs primarily in the enclosed volume.

To make this spatial power distribution more transparent, we briefly consider for simplicity an axially symmetric case of a long conducting cylinder with radius $R$ and electrical conductivity $\sigma$, with magnetic fields directed along the cylinder axis $\hat{\boldsymbol{z}}$, so that we can use cylindrical coordinates and focus on the respective components $B = B_z$, $j_{ind} = j_{ind,\varphi}$, $E = E_\varphi$. The local Joule heating power density is

$$p_J(r,t) = j_{ind}(r,t)^2/\sigma. \tag{4}$$

Using the Maxwell equations $-\partial B(r,t)/\partial t = \partial/\partial r[rE(r,t)]/r$, $H_{ind}(r,t) = \int_r^R j_{ind}(r',t)dr'$ and Ohm's law $j_{ind}(r,t) = \sigma E(r,t)$, the local power density $p_{ind}$ becomes with equation (3)

$$p_{ind}(r,t) = \left(-\partial\left[r j_{ind}(r,t)\right]/\partial r\right) \int_r^R j_{ind}(r',t)dr' /(r\sigma). \tag{5}$$

Obviously, the $p_J(r,t)$ and $-p_{ind}(r,t)$, expressed by $j_{ind}(r,t)$, are not equal, so that the Joule heating profile is different from the distribution of $p_{ind}(r,t)$. However, we show in Section 2 of the Supplementary material that for arbitrary geometries and $\sigma(\boldsymbol{r})$, the corresponding quantities $P_J(t)$ and $P_{ind}(t)$ obtained by integration over the whole sample volume, exactly cancel out at any time $t$, thus ensuring global energy conservation.

We now return to the case of a superconductor in the Meissner state, which is the main subject of this article. Neglecting all extrinsic effects, the role of $\boldsymbol{j_{ind}}(\boldsymbol{r}, t)$ induced by $\partial \boldsymbol{B}/\partial t$ is taken by the normal current $\boldsymbol{j_n}(\boldsymbol{r}, t)$ in the spirit of equation (1b), which represents both the source of Joule heating and magnetocaloric cooling according to equations (4) and (5). If we have exact solutions of equation (1b) for $\boldsymbol{B}(\boldsymbol{r}, t)$ which may differ from those of (1a), or of any other equation that correctly describes the time dependence of the Meissner effect but also leads to normal currents determined by Faraday's law of induction, the same formalism and conclusions from above apply without obvious restrictions. This is one of the central results of our article. We emphasize here that the dissipationless supercurrent $\boldsymbol{j_s}(\boldsymbol{r}, t)$ is not caused by Faraday's law, but is, due to its quantum mechanical origin, given by its proportionality to $-\boldsymbol{A}(\boldsymbol{r}, t)$. Its possible time dependence is reflected in that of $\boldsymbol{B}(\boldsymbol{r}, t)$, which in turn provides the driving force for $\boldsymbol{j_n}(\boldsymbol{r}, t)$. A corresponding analogue to equation (3) for $\boldsymbol{j_s}$ would represent the variation in magnetic energy density $u_{mag} = \boldsymbol{B}^2/(2\mu_0)$, with $\partial u_{mag}/\partial t = \boldsymbol{B} \cdot (\partial \boldsymbol{B}/\partial t)/\mu_0$.

## 4. Numerical estimates

### *4.1. Magnetic-field distribution*

We first briefly estimate the order of magnitude of $\boldsymbol{j_n}$ that we can expect in real superconductors and examine its effect on the magnetic-field profile. We suppose the validity of equation (1b) and that we are dealing with an idealized type-I superconductor, but with material parameters similar to those of niobium with $T_c$= 9 K [27,28] and an effective penetration depth $\lambda_0$ = 47 nm at $T$ = 0 [29] with a two-fluid-like temperature dependence, $\lambda(T)^{-2} = \lambda_0^{-2}(1 - [T/T_c]^4)$. Since a possible magnetic field dependence of $\lambda$ in type I superconductors is known to be very weak [30], we can neglect it for the present analysis. We also assume that the critical magnetic field varies as $B_c(T) = B_c(0)(1 - [T/T_c]^2)$, with $B_c(0) \approx 0.2$ T [28,31] (see Fig. 1b). For simplicity, we restrict our estimates for the moment to thermodynamic changes of state due to temperature variations that occur strictly within the Meissner phase, ignoring possible complications that may arise for a situation where $T_c$ is crossed, such as the consumption or the release of a latent heat and an associated discontinuous jump in physical quantities. To account for a possible $T$-dependence of $\sigma$ below $T_c$, we assume that it obeys a Drude-type law where $\sigma(T)$ is proportional to $n_n$, i.e., $\propto (T/T_c)^4$, and to a material-specific scattering time $\tau(T)$, the $T$-dependence of which we take from corresponding resistivity data where superconductivity had been quenched by a magnetic field (see Supplementary material and ref. [32]). For consistency, we normalize it to a realistic value at $T_c$, $\sigma(T_c) = \sigma_n = 1.8\text{x}10^9$ $(\Omega\text{m})^{-1}$, which we estimate from the thermal conductivity $\kappa \approx 400$ W/(Km) given in ref. [33] at $T_c$ using the Wiedemann-Franz law, $\sigma_n = \kappa/(LT)$, with $L = 2.44\text{x}10^{-8}$ $\text{W}\Omega\text{K}^{-2}$. However, as we shall see below, significant measurable thermal effects are only expected to occur close to $T_c$, so that the knowledge of the exact temperature dependence below $T_c$ is not crucial in this respect.

During field cooling a corresponding infinitely long superconducting cylinder with radius $R = 6\lambda_0$ in a $B_0 = \mu_0 H_0$ well below the critical field $B_0 \ll B_c(0)$ at a realistic cooling rate $dT/dt = -0.05$ K/s across the critical temperature (the weak field dependence of which we ignore for this consideration), the $B_z(r,t)/B_0$ as obtained by numerically solving equation (1b) (see Section 4 in the Supplementary material) is virtually indistinguishable from the corresponding static solution given by equation (1a) and plotted in Fig. 2a. Pronounced deviations would only occur in a regime where $\boldsymbol{j_n}$ becomes comparable to the order of magnitude of $\boldsymbol{j_s}$. The dimensionless, $T$-independent parameter $\epsilon = (\mu_0 \sigma_n \lambda(0)^2 / T_c)(dT/dt)$ is related to the ratio of the leading prefactors of $\boldsymbol{j_n}(\boldsymbol{r})$ and $\boldsymbol{j_s}(\boldsymbol{r})$ as derived from the respective analytical solution of equation (1a) and given in equations (12) and (13) of the Supplementary material. In our model superconductor, $\epsilon \approx 3\text{x}10^{-14}$ only, which is a typical order of magnitude for existing superconductors. We have also tentatively assumed extremely high hypothetical $\sigma_n$ values between $2.6\text{x}10^{20}$ $(\Omega\text{m})^{-1}$ ($\epsilon \approx 0.004$) and $2.6\text{x}10^{23}$ $(\Omega\text{m})^{-1}$ ($\epsilon \approx 4$) and plotted the resulting $B_z(r,t)/B_0$ in Figs. 2b-2f for comparison. However, the crossover criterion $\epsilon \approx 1$ in Fig. 2d is by many orders of magnitudes unreachable for existing superconductors under realistic experimental field-cooling or warming conditions, except perhaps in an extremely narrow region around $T_c$, so that possible effects of a finite $\boldsymbol{j_n}$ on $\boldsymbol{B}(\boldsymbol{r},t)$ are probably not directly measurable. We can nevertheless conclude from this important consideration that the solutions of equation (1a) serve as an excellent approximation for the quantitative modelling of slowly varying magnetic fields and currents in real superconducting systems, and we will therefore use the corresponding analytic solutions in the following.

*4.2. Thermal effects*

On this basis, we provide in Section 3 of the Supplementary material explicit formulae for $p_J(r,t)$ and $p_{ind}(r,t)$ for a cylinder geometry and in the realistic limit $\epsilon \ll 1$, where not only $\boldsymbol{B}(\boldsymbol{r})$ but also all relevant current densities are virtually indistinguishable from the static solutions associated with equation (1a). The magnitude of both power densities is determined by the quantity $\sigma(B_0 \frac{\partial\lambda}{\partial T}\frac{dT}{dt})^2$, while their spatial distribution is clearly different (see Section 3 of the Supplementary material). We now consider a cooling with $dT/dt$ = –0.05 K/s in external magnetic fields $B_0$ at different temperatures $T$ just below the respective $T_c(B_0)$, where the superconductor is in the Meissner phase and can be assumed to be in quasistatic equilibrium. In Figs. 3a-3c, we show the resulting $p_J(r,t)$ and $p_{ind}(r,t)$ data for our model superconductor and for different particle sizes. The regions of pronounced local heat generation by $p_J(r,t)$ where the normal currents $\boldsymbol{j_n}(\boldsymbol{r},t)$ are strongest are always qualitatively closer to the edge of the superconductor than the regions where the cooling power $p_{ind}(r,t)$ dominates. We illustrate this in Fig. 3d, where the relevant length scales are plotted in dimensionless units that neither depend on the choice of material parameters nor on a particular temperature protocol. For small particles or close to $T_c$ where $R/\lambda(T)$ is small, the heated zone with $p_J > |p_{ind}|$ covers a significant area of the cross section, while it is squeezed towards the edge with increasing $R/\lambda(T)$, i.e., for large sample dimensions or low temperatures, respectively. The variations in the integrated total Joule heating and magnetocaloric cooling powers per unit length of the cylinder as obtained from equation (17) in the Supplementary material are shown in Fig. 3e for different temperatures $T$ and cylinder radii $R$.
The spatial separation of heated and cooled zones, as given by equations (4), (5) and shown in Figs. 3a-3d, should result in the formation of certain temperature gradients, the extent and magnitude of which depend on the rate of heat diffusion. To obtain the resulting temperature distributions, we have to solve the thermal diffusion equation $\partial T(\boldsymbol{r},t)/\partial t - \kappa\Delta T(\boldsymbol{r},t)/C = (p_J(r,t) + p_{ind}(r,t))/C$ (where $\kappa$ is the thermal conductivity and $C$ is the heat capacity per volume), which again turns out to be only numerically possible (see Section 4 in the Supplementary material). The expected local deviations from the mean temperature obtained with $\kappa \approx 400$ W/(Km) [33] and $C \approx 2.7\text{x}10^4$ J/(Km$^3$) [34] are plotted in the insets of Figs. 3a-3c and show that the temperature gradients increase with increasing cylinder radius $R$ and increasing $T \lesssim T_c(B_0)$ towards $T_c(0)$ even though the applied magnetic fields $B_0$ decrease. This is a consequence of the fact that $\partial\lambda/\partial T$ grows faster towards $T_c(0)$ than $B_c(T)$ decreases. Even under the most favorable assumptions, these temperature gradients are extremely small. They may therefore have gone unnoticed until now and may even be beyond the reach of current experiments. However, this seemingly discouraging statement is only true for thermodynamic changes of state well within the Meissner state at temperatures below the corresponding $T_c(B_0)$. At the critical temperature itself, on the other hand, the first-order nature of the transition to superconductivity in a magnetic field in type-I superconductors must lead to a sharp jump in $\lambda(T)$ between infinity and a finite value with a singular $\partial\lambda/\partial T$, resulting in a virtually discontinuous flux expulsion on cooling across $T_c(B_0)$. The associated induced normal currents are therefore expected to be by orders of magnitude stronger than those occurring during slow temperature variations deep in the Meissner phase as considered above, especially if $B_0$ is chosen to be a significant fraction of $B_c(0)$. As a result, both the Joule heating and magnetocaloric cooling powers should in principle become very large there and could indeed lead to measurable thermal effects, thus providing an opportunity to prove or disprove the concepts developed here. These temperature gradients might be accessible, for example, using a simplified form of ac calorimetry-type techniques [35,36]. In such experiments, the average temperature of a sample is made to oscillate by means of an external heater. Both $p_J$ and $p_{ind}$ are proportional to $(dT/dt)^2$ and thus to the square of the chosen oscillation frequency. In this way, the maximum $dT/dt$ of corresponding temperature oscillations can be made substantially larger than in heating or cooling experiments with a constant $dT/dt$, and the temperature window of interest can be probed any number of times. Such an experiment should be carried out on wide samples, since the total Joule heat increases with $R$ (see Fig. 3e). Two ends of a pair of tiny differential thermocouples could be attached, one at the edge and one toward the center of the superconductor where, according to Figs. 3a-3d, the difference in the extreme values of $p_J$ and $p_{ind}$ is expected to be largest.

The behavior right at the phase transition is very difficult to analyze and model quantitatively, however, because it depends on uncontrollable details of the phase transition, including the possible occurrence of geometry-dependent inhomogeneous intermediate states, supercooling or superheating phenomena, and temperature inhomogeneities resulting from a non-uniform latent-heat release or consumption in different superconducting domains. Therefore, for this case, we cannot provide a reliable estimate of the magnitude of the expected temperature inhomogeneities at this time.

## 5. Conclusions

In summary, our results have shown that the presence of unpaired charge carriers in a superconductor should lead to deviations of the magnetic-field distribution from the textbook Meissner profile during dynamic thermodynamic changes of state. The associated normal currents that are induced by the variations in the magnetic-flux density must also generate Joule heating and, at the same time, magnetocaloric cooling in spatially distinct regions of the superconductor, while always maintaining the overall energy balance and, ultimately, ensuring reversibility in terms of energy. It remains to be seen whether the energy balance discussed here has implications for other areas of research on superconductors, where the dissipation of induced normal currents plays a prominent role. The physical concept used here to treat the energy balance in a conducting magnetic material is not restricted to superconductors but can be directly applied to other conducting materials with a temperature-dependent magnetic susceptibility, such as metallic ferromagnets, or to describe the thermodynamics of thermomagnetic generators.

**Data availability statement** All data that support the findings of this study are included within the article (and any supplementary files). The corresponding Maple worksheets used for the calculations are available on reasonable request.

**Acknowledgments** We thank to S.M. Anlage, V.G. Kogan and R. Prozorov for useful comments and discussions, and to A. Drake for preparing graphic animations.

## Figures

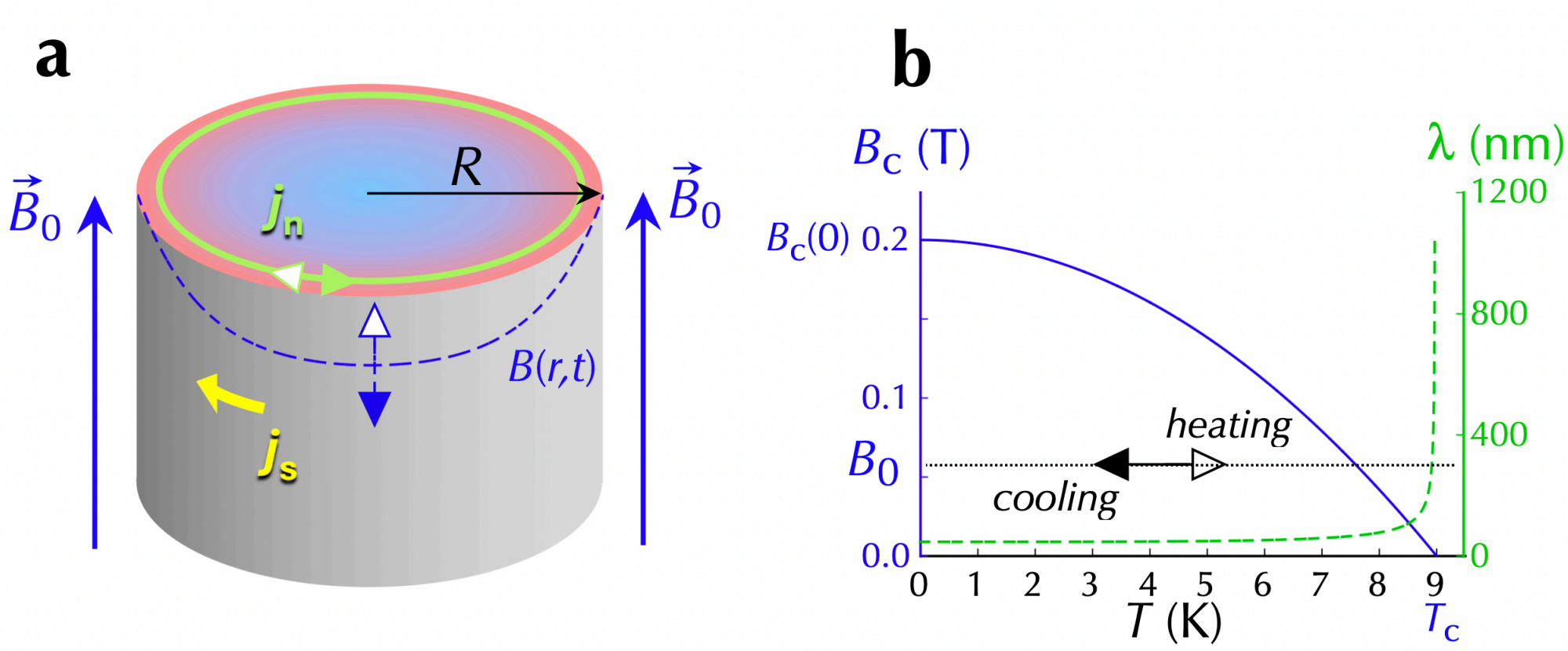

**Figure 1.** Sketch of the experimental situation discussed in the present article. (a) A long cylindrical type-I superconductor is placed in an external magnetic field $\boldsymbol{B}_0$ directed parallel to the cylinder axis. Below the critical temperature $T_c$ and for $\boldsymbol{B}_0$ smaller than the critical field $\boldsymbol{B}_c(T)$, the material is in the Meissner phase, with a typical Meissner profile $B(r)$ (sketched with a dashed line) due to the shielding supercurrent $j_s$. When the temperature is changed (cooling symbolized by filled arrows, heating by open arrows), $B(r,t)$ becomes time-dependent and induces normal currents $j_n$ due to unpaired charge carriers, whose direction depends on the chosen temperature protocol. These currents always cause Joule heating near the edge (red color) and magnetocaloric cooling (blue color) toward in the inside of the cylinder. (b) Critical field $B_c(T)$ and magnetic penetration depth $\lambda(T)$ of the chosen type-I model superconductor.

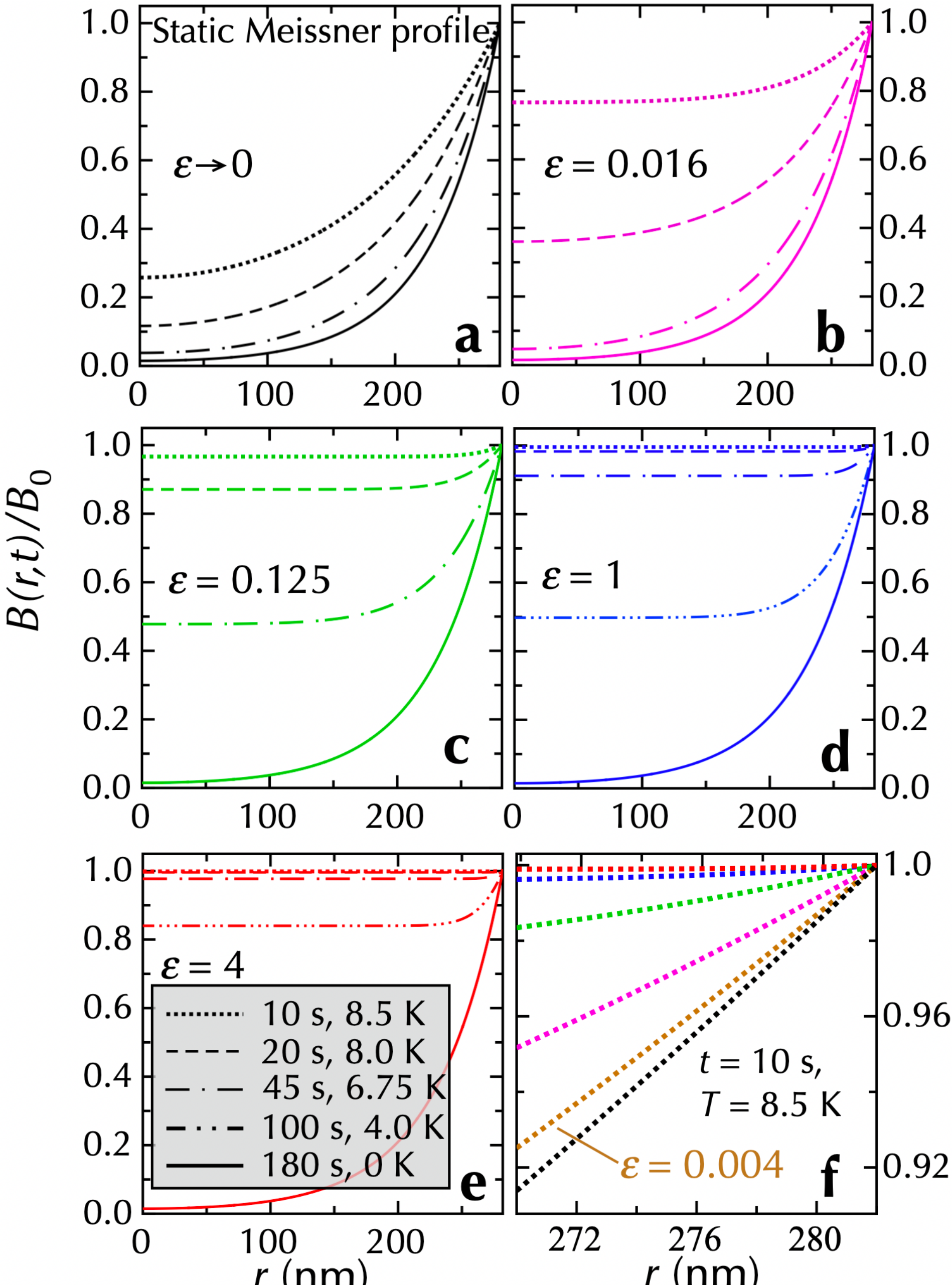

**Figure 2.** Magnetic-field distributions $B(r,t)/B_0$ obtained by numerically solving equation (1b) for a type-I superconducting cylinder with Radius $R = 6\lambda_0 = 0.282$ μm and the model parameters as described in the text, at times $t$ = 10 s, 20 s, 45 s, 100 s (for $\epsilon = 1$ and 4 only) and 180 s, respectively, after entering the superconducting state at $T_c \approx 9$ K with $\mathrm{d}T/\mathrm{d}t = -0.05$ K/s in a weak magnetic field $B_0 \ll B_c(0)$. (a) For the chosen model superconductor with $\epsilon \approx 3\text{x}10^{-14}$, $B(r,t)/B_0$ is virtually indistinguishable from the static Meissner profile, $\epsilon \rightarrow 0$. (b)-(e) Corresponding data for $\epsilon$ around unity. (f) Expanded view of the data near the edge of the cylinder for $t$ = 10 s. All hypothetical dynamic profiles for $\epsilon > 0$ would relax to the corresponding static Meissner profiles ($\epsilon = 0$) on a time scale $\mu_0\sigma\lambda^2$ [10] proportional to $\epsilon$ after a thermodynamic change of state has stopped.

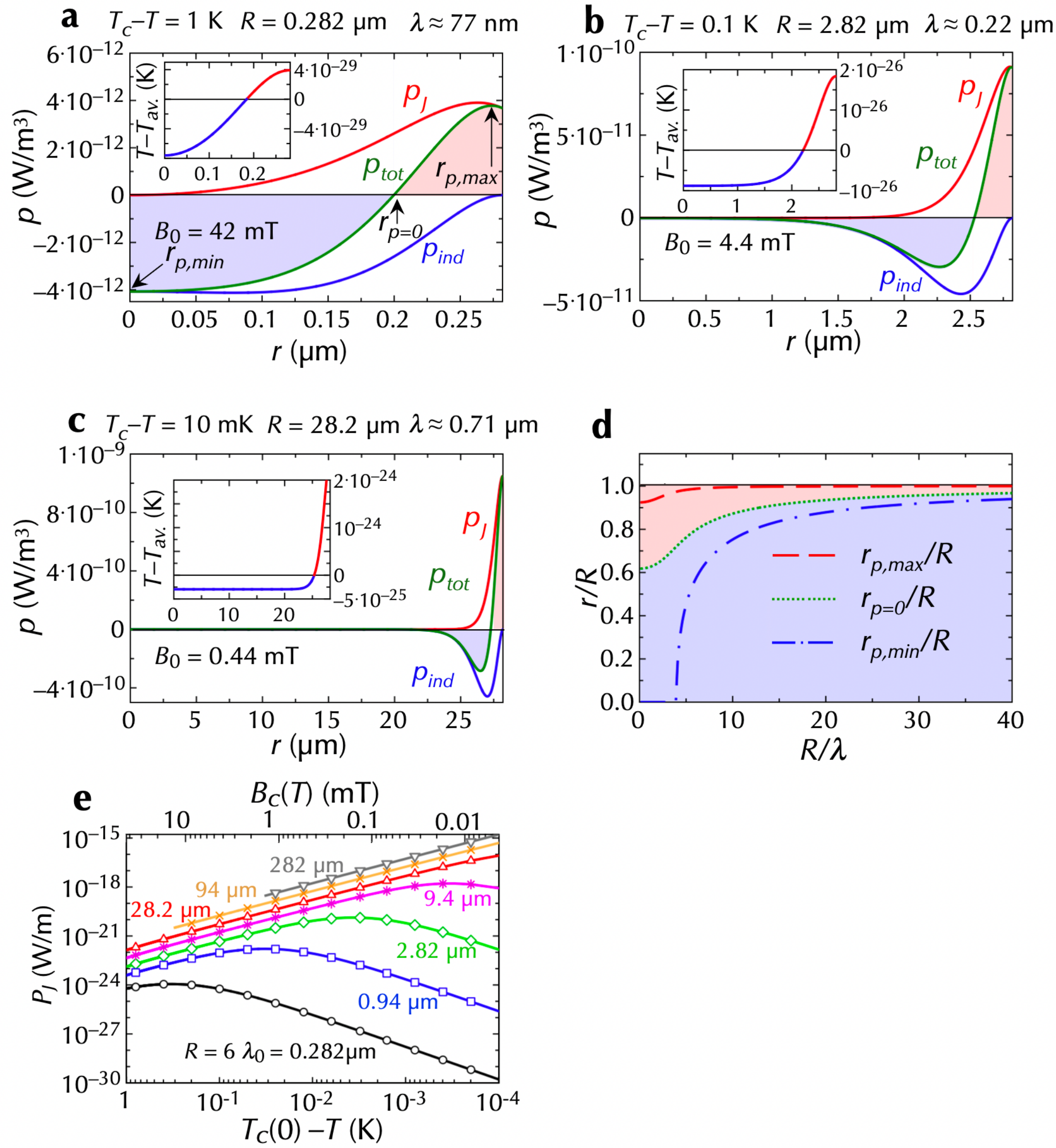

**Figure 3.** Local and integrated power densities for the model type-I superconductor as described in the text. (a)-(c) Local power densities $p_J(r,t)$, $p_{ind}(r,t)$ and $p_{tot}(r,t) = p_J(r,t) + p_{ind}(r,t)$ for $R = 6\lambda_0$, $60\lambda_0$, and $600\lambda_0$ (0.282 μm, 2.82μm, and 28.2 μm) respectively, in different external magnetic fields $B_0$ and at temperatures $T$ just below the corresponding $T_c(B_0)$ in the Meissner phase, where the temperature is allowed to vary with –0.05 K/s. The insets show the resulting local deviations from the mean temperature, $T_{av}$. We have marked $r_{p,max}$, $r_{p=0}$, and $r_{p,min}$ in Fig. 3a, where the total local power density $p_{tot}$ reaches its maximum, zero, or minimum value, respectively. (d) Corresponding values in dimensionless units as functions of $R/\lambda(T)$. In all these figures, the light red and light blue coloring schemes symbolize local heating and local cooling, respectively. (e) Integrated power $P_J(t) = -P_{ind}(t)$ per unit length of the cylinder for different values of $R$ near the critical temperature $T_c(0)$, illustrating the increasing trend of $P_J(t)$ with increasing $R$ and $T$.

**Supplementary material: Energy conservation and reversibility during thermodynamic changes of state in superconductors: Joule heat vs. magnetocaloric cooling**

Andreas Schilling* ORCID 0000-0002-3898-2498
Dept. of Physics, University of Zürich, 8057 Zürich, Switzerland

*Corresponding author. *Email address:* schilling@physik.uzh.ch (A. Schilling)

## 1. List of cases with sign combinations of $\frac{\partial \boldsymbol{B}}{\partial t}$ and $\boldsymbol{H_{ind}}$ implying that $\boldsymbol{p_{ind}} < \boldsymbol{0}$.

| $\partial M/\partial T$ | $dT/dt$ | $\partial B/\partial t$ | $H_{ind}$ | $p_{ind} = H_{ind}\partial B/\partial t$ |
|---|---|---|---|---|
| $>0$ | $>0$ | $>0$ | $<0$ | $<\mathbf{0}$ |
| $>0$ | $<0$ | $<0$ | $>0$ | $<\mathbf{0}$ |
| $<0$ | $>0$ | $<0$ | $>0$ | $<\mathbf{0}$ |
| $<0$ | $<0$ | $>0$ | $<0$ | $<\mathbf{0}$ |

For a superconductor in the Meissner state, $\partial M/\partial T > 0$.

## 2. Proof of $\boldsymbol{P_{ind}} + \boldsymbol{P_J} = \boldsymbol{0}$ for arbitrary geometries and $\boldsymbol{\sigma(r)}$

The local Joule heating power due to the induced currents $\boldsymbol{j_{ind}}(\boldsymbol{r}, t)$ is

$$p_J(\boldsymbol{r}, t) = \frac{\boldsymbol{j_{ind}}(\boldsymbol{r},t)^2}{\sigma(\boldsymbol{r})}, \tag{1}$$

while the magnetic field produced by them is given by the Maxwell equation

$$\boldsymbol{\nabla} \times \boldsymbol{H_{ind}}(\boldsymbol{r}, t) = \boldsymbol{j_{ind}}(\boldsymbol{r}, t). \tag{2}$$

As outlined in equation (3) of the main article, the power density $p_{ind}(\boldsymbol{r})$ is given in general by

$$p_{ind}(\boldsymbol{r}, t) = \boldsymbol{H_{ind}}(\boldsymbol{r}, t) \cdot \frac{\partial \boldsymbol{B}(\boldsymbol{r},t)}{\partial t}. \tag{3}$$

Because $\boldsymbol{j_{ind}}$ is finite at the surface of an extended finite conducting body, the corresponding boundary condition for $\boldsymbol{H_{ind}}(\boldsymbol{r}, t)$ is $\boldsymbol{H_{ind,surf}} = \boldsymbol{0}$ so that the total magnetic field $\boldsymbol{H} = \boldsymbol{H_0} + \boldsymbol{H_{ind}}$ remains continuous and $p_{ind} = 0$ at the surface. With the Maxwell equation

$$\frac{\partial \boldsymbol{B}(\boldsymbol{r},t)}{\partial t} = -\boldsymbol{\nabla} \times \boldsymbol{E}(\boldsymbol{r}, t) = -\boldsymbol{\nabla} \times \frac{\boldsymbol{j_{ind}}(\boldsymbol{r},t)}{\sigma(\boldsymbol{r})}, \tag{4}$$

we have

$$p_{ind}(\boldsymbol{r},t) = -\boldsymbol{H_{ind}}(\boldsymbol{r},t)\cdot\left\{\boldsymbol{\nabla}\times\frac{j_{ind}(\boldsymbol{r},t)}{\sigma(\boldsymbol{r})}\right\}. \quad (5)$$

The total power density $P_{ind} + P_J$, integrated over the whole sample volume then becomes

$$P_{ind} + P_J =$$

$$\int_{vol} d\boldsymbol{r}\left[p_{ind}(\boldsymbol{r},t) + p_J(\boldsymbol{r},t)\right] = \int_{vol} d\boldsymbol{r}\left[-\boldsymbol{H_{ind}}(\boldsymbol{r},t)\cdot\left\{\boldsymbol{\nabla}\times\frac{j_{ind}(\boldsymbol{r},t)}{\sigma(\boldsymbol{r})}\right\} + \frac{j_{ind}(\boldsymbol{r},t)^2}{\sigma(\boldsymbol{r})}\right]. \quad (6)$$

With the use of the vector identity

$$\boldsymbol{\nabla}\cdot(\boldsymbol{H_{ind}}(\boldsymbol{r},t)\times\frac{j_{ind}(\boldsymbol{r},t)}{\sigma(\boldsymbol{r})}) = \left(\boldsymbol{\nabla}\times\boldsymbol{H_{ind}}(\boldsymbol{r},t)\right)\cdot\frac{j_{ind}(\boldsymbol{r},t)}{\sigma(\boldsymbol{r})} - \boldsymbol{H_{ind}}(\boldsymbol{r},t)\cdot\left\{\boldsymbol{\nabla}\times(\frac{j_{ind}(\boldsymbol{r},t)}{\sigma(\boldsymbol{r})})\right\}, \quad (7)$$

the boundary condition $\boldsymbol{H_{ind,surf}} = \boldsymbol{0}$, and equation (2), we obtain

$$P_{ind} + P_J = \int_{vol} d\boldsymbol{r}\left[\boldsymbol{\nabla}\cdot(\boldsymbol{H_{ind}}(\boldsymbol{r},t)\times\frac{j_{ind}(\boldsymbol{r},t)}{\sigma(\boldsymbol{r})}) - \left(\boldsymbol{\nabla}\times\boldsymbol{H_{ind}}(\boldsymbol{r},t)\right)\cdot\frac{j_{ind}(\boldsymbol{r},t)}{\sigma(\boldsymbol{r})} + \frac{j_{ind}(\boldsymbol{r},t)^2}{\sigma(\boldsymbol{r})}\right] \quad (8)$$

$$= \int_{vol} d\boldsymbol{r}\left[\boldsymbol{\nabla}\cdot(\boldsymbol{H_{ind}}(\boldsymbol{r},t)\times\frac{j_{ind}(\boldsymbol{r},t)}{\sigma(\boldsymbol{r})})\right] = \int_{surface} d\boldsymbol{n}\cdot(\boldsymbol{H_{ind}}(\boldsymbol{r},t)\times\frac{j_{ind}(\boldsymbol{r},t)}{\sigma(\boldsymbol{r})}) = 0. \quad (9)$$

## 3. Explicit expressions for cylindrical superconductors in the Meissner state

Explicit expressions for thermodynamic changes of state of a superconductor in the Meissner state and in a constant external magnetic field $\boldsymbol{H_0}$ can be obtained in the limit $\partial T/\partial t \ll T_c/(\mu_0\sigma_n\lambda_0^2)$ $(\epsilon \ll 1)$ which is always fulfilled for existing superconductors in a realistic experiment (see main article text). The resulting magnetic-field distribution is then virtually indistinguishable from that of a static London-like profile, which is for a long cylinder with radius $R$ in cylindrical coordinates

$$B(r,t) = \mu_0 H_0 \frac{I_0\left(\frac{r}{\lambda}\right)}{I_0\left(\frac{R}{\lambda}\right)}. \quad (10)$$

We are using here and below the modified Bessel functions of the first kind $I_0(x)$ and $I_1(x)$, $\lambda(T(t))$ the temperature (and therefore time dependent) effective penetration depth, and $\sigma$ the normal-state electrical conductivity that we assume to be spatially constant. With

$$\frac{\partial B(r,t)}{\partial t} = \mu_0 H_0 \frac{\partial\lambda}{\partial T}\frac{dT}{dt}\frac{-rI_0\left(\frac{R}{\lambda}\right)I_1\left(\frac{r}{\lambda}\right) + RI_0\left(\frac{r}{\lambda}\right)I_1\left(\frac{R}{\lambda}\right)}{\lambda^2 I_0\left(\frac{R}{\lambda}\right)^2} \quad (11)$$

and $rj_n(r,t) = -\sigma\int_0^r r'\frac{\partial B(r',t)}{\partial t}dr'$ with $j_n(0,t) = 0$ we obtain

$$j_n(r,t) = -\mu_0 H_0 \sigma \frac{\partial\lambda}{\partial T}\frac{dT}{dt}\frac{\left[-rI_0\left(\frac{r}{\lambda}\right)+2\lambda I_1\left(\frac{r}{\lambda}\right)\right]I_0\left(\frac{R}{\lambda}\right)+RI_1\left(\frac{R}{\lambda}\right)I_1\left(\frac{r}{\lambda}\right)}{\lambda I_0\left(\frac{R}{\lambda}\right)^2}, \tag{12}$$

whereas the supercurrent density is

$$j_s(r) = j_{tot}(x) - j_n(r,t) \approx j_{tot}(x) = -\frac{H_0}{\lambda}\frac{I_1\left(\frac{r}{\lambda}\right)}{I_0\left(\frac{R}{\lambda}\right)}. \tag{13}$$

The resulting local Joule power density is

$$p_J(r,t) = \frac{j_n(r,t)^2}{\sigma}. \tag{14}$$

The $H_{ind}(r) = \int_r^R j_n(r')dr'$ becomes

$$H_{ind}(r,t) = \mu_0 H_0 \sigma \frac{\partial\lambda}{\partial T}\frac{dT}{dt}\left(\frac{2\lambda I_0\left(\frac{r}{\lambda}\right)I_0\left(\frac{R}{\lambda}\right)+RI_0\left(\frac{r}{\lambda}\right)I_1\left(\frac{R}{\lambda}\right)-rI_0\left(\frac{R}{\lambda}\right)I_1\left(\frac{r}{\lambda}\right)}{I_0\left(\frac{R}{\lambda}\right)^2} - 2\lambda\right), \tag{15}$$

which can be used to calculate the local magnetocaloric cooling power with equation (11),

$$p_{ind}(r,t) = H_{ind}(r,t)\frac{\partial B(r,t)}{\partial t}. \tag{16}$$

The total Joule power $P_J = -P_{ind}$, integrated over the cross section of the cylinder is with $A_0 = I_0\left(\frac{R}{\lambda}\right)$ and $A_1 = I_1\left(\frac{R}{\lambda}\right)$

$$P_J(t) = \frac{\pi R\sigma\left(\mu_0 H_0 \frac{\partial\lambda}{\partial T}\frac{dT}{dt}\right)^2}{3\lambda^2 A_0^4} \times \tag{17}$$

$$[A_0^4(R^3 - 12\lambda^2 R) + A_0^3 A_1(24\lambda^3 - 10\lambda R^2) + A_0^2 A_1^2(22\lambda^2 R - 4R^3) + 12A_0 A_1^3 \lambda R^2 + 3A_1^4 R^3].$$

## 4. Methods

### Modelling the dc electrical conductivity of the normal carriers below $T_c$

The temperature dependence of the dc electrical conductivity $\sigma$ of the normal carriers inserted in equation (1b) was treated as follows: We assumed that $\sigma(T)$ obeys a Drude-type law where $\sigma$ is proportional to $n_n$, i.e., $\propto (T/T_c)^4$, and to a material-specific scattering time $\tau(T)$, the $T$-dependence of which we took from corresponding resistivity data of niobium reported in ref. [1], where superconductivity was quenched by a magnetic field. This leads to

$$\frac{\sigma(T)}{\sigma_n(T_c)} \approx \frac{(T/T_c)^4}{0.1829 + 3.950\text{x}10^{-2}(T/T_c)^2 + 0.7776(T/T_c)^3} \tag{18}$$

and is roughly proportional to $(T/T_c)^2$. In the numerical procedure, equation (18) was used, however.

**Numerical procedures**

To numerically solve equation (1b) and the thermal diffusion equation mentioned in the main text, we used the software Maple 2024 by Maplesoft, Waterloo, Canada. The thermal gradients obeying the heat-diffusion equation, $\partial T(\boldsymbol{r},t)/\partial t - \kappa \Delta T(\boldsymbol{r},t)/C = p_{tot}(\boldsymbol{r},t)/C$ (with $\kappa$ the thermal conductivity and $C$ the heat capacity per volume) could alternatively be estimated by using the fact that the thermal diffusivity $\kappa/C$ of metallic superconductors around $T_c$ is sufficiently large ($\kappa/C \approx 0.02$ m$^2$/s) [2,3] so that the temperature compensation across the steepest power gradient along $p_{tot}(\boldsymbol{r},t)$ is fast enough to approach quasi-stationary conditions for reasonably slow changes of the thermodynamic state. Then, the $\partial T(r,t)/\partial t$ becomes negligibly small compared to the other terms, and the Poisson equation $\Delta T(\boldsymbol{r},t) \approx - p_{tot}(\boldsymbol{r},t)/\kappa$ holds over extended parts of the cylinder volume, the solutions of which also provide useful estimates of the temperature distribution. The $p_{tot}(r,t) = p_J(r,t) + p_{ind}(r,t)$ for the limiting case $\epsilon \ll 0$ can been included here, for which analytical solutions are given in the equations (11) – (16).